\documentclass[twocolumn,showpacs,preprintnumbers,amsmath,amssymb]{revtex4}
\usepackage{amsfonts}
\usepackage{amsmath}
\usepackage{graphicx}
\usepackage{dcolumn}
\usepackage{bm}
\usepackage{overpic}
\usepackage{booktabs}

\begin{document}
\title{Unevenness of Loop Location in Complex Networks}
\author{An Zeng, Yanqing Hu, Zengru Di\footnote{zdi@bnu.edu.cn}}
 \affiliation{Department of Systems Science, School of Management and Center for Complexity
 Research, Beijing Normal University, Beijing 100875, China}

\date{\today}

\begin{abstract}
The loop structure plays an important role in many aspects of
complex networks and attracts much attention. Among the previous
works, Bianconi et al find that real networks often have fewer short
loops as compared to random models. In this paper, we focus on the
uneven location of loops which makes some parts of the network rich
while some other parts sparse in loops. We propose a node removing
process to analyze the unevenness and find rich loop cores can exist
in many real networks such as neural networks and food web networks.
Finally, an index is presented to quantify the unevenness of loop
location in complex networks.


\end{abstract}



\maketitle

\textbf{Introduction.} In the last decade, the researches on complex
networks have rapidly developed. At the same time, the loop
structure has attracted much attention. Loops are very important in
complex networks. They can not only characterize the detail
structure of networks but also relate to the structural
correlations, motifs, robustness and redundancy of pathways, and
affect some dynamical as well as equilibrium critical phenomena of
the networks[1,2]. Recently, to avoid the effect of the loop
structure, some researchers even study the acyclic networks[3,4].

For the self-avoiding loop, researchers focus mainly on two aspects:
the total number of loops and the the dynamic effect of the loop
structure. In the former case, many counting methods have been
proposed[5-11]. In undirected networks, short loops can be
exactly counted in terms of powers of the adjacency matrix[8]. This
method can not deal with long loops, because the counting
equation will become very complicated when facing long loops. In
directed networks, short loops can be estimated by using
$N_{L}\simeq \frac{1}{L} Tr A^{L}$ while long loops can be
calculated by the entropy[9,10]. Moreover, some researchers analyze
statistics of loops with different length $L$. They use the Monte
Carlo sampling method to get the frequency and find that the loop
number is sharply peaked around a characteristic loop length
$L^{*}$. Also, they use $L^{*}$ and the relevant index to
characterize the networks[11]. On the other hand, the dynamic effect
of the loop structure has been studied frequently. It has been
pointed out that the loop structure is related to the activity in
neural networks such as self-sustained activities[12,13,14] and
synchronization[4,15,16]. Specifically, the self-sustained activity
can not survive without the loop structure and the synchronization
will be weakened when emerging a dominant loop in the network.
What's more, a scaling behavior of loops is used to explain some
critical phenomenon in percolation[17] and loop number is also used
as a ranking method to quantify the role of both nodes and
links[18].

However, many problems about loops still remain unnoticed. In the
ref.[10], Bianconi, Gulbahce and Motter find that many real networks
have fewer loops than the counterpart random networks which are a
kind of random networks with the same number of in- and out-links in
each node as the real networks. Actually, the loops number in
different parts of a network varies according to the function of the
regions. For example, the feed-forward part of the neural networks
are sparse of loops[3] while other parts in the brain need loops to
carry out self-sustained oscillation for precessing
information[13,14]. For the food web networks, the loops number in
the metazoan part are relatively small while there are many short
loops among the microorganisms, called microbial loops, for fixed
carbon repacking and recovery path of ecosystem[19]. Obviously,
loops locate unevenly in many real networks. Some communities of
these networks will be rich in loops while loops will be sparse in
other parts. This leads us to an interesting question: what is the
detail organization of loops location like in the networks? In this
paper, we focus on the unevenness of loops location. We first study
the distribution of loops on single nodes. Then we analyze the rich
loop core phenomenon of uneven loops location by a node removing
process in some real networks. Finally, we propose an index to
measure the unevenness.

\textbf{Heterogenous distribution of loops on single nodes.} In the
first step, we should study the loops on each single node to help us
understand how the loops locate in the network. For a given
network with size $N$, if we want to obtain how many loops passing through a
specific node, we can simply remove the node from the network and
count how many loops decreases, the decrement is the number of loops
on this node. As $N_{L}$ is the loops number with the length $L$ of
a network, we denote the $\hat{N}_{L}(i)$ as the number of loops with
length $L$ in the network after the node $i$ is removed. So the node
has $C_{L}(i)=N_{L}-\hat{N}_{L}(i)$ loops with length $L$ passing
through. The number of short loops in directed network can be
expressed in terms of powers of the adjacency matrix. In particular,
$N_{L}\simeq \frac{1}{L} Tr A^{L}$,
 provided that $\kappa\equiv
max_{i}\sum_{j}\sum^{'}_{m}(\begin{array}{c}l\\m\end{array})|\lambda^{-m}_{j}P_{ij}P^{-1}_{j+m,i}|\ll1$[9].
Because $C_{L}(i)=N_{L}-\hat{N}_{L}(i)$, we can count the short loops
on single nodes in any networks.

Here, we focus on the distribution of $C_{L}(i)$ in different
networks. We can compare $C_{L}^{r}(i)$ of real networks with
$C_{L}^{c}(i)$ of the counterpart random networks and $C_{L}^{e}(i)$
of the corresponding ER random networks. The counterpart random
networks are a kind of uncorrelated random networks with the same
number of in- and out-links in each node as the real networks. The
corresponding ER random networks is given with the same size and the
total number of links as the real networks. Of course,
$C_{L}^{r}(i)$, $C_{L}^{c}(i)$ and $C_{L}^{e}(i)$ can be obtained by
$C_{L}(i)=N_{L}-\hat{N}_{L}(i)$. Actually, the expected value of
$C_{L}^{c}(i)$ can be gained by the formula based on the degree
sequence. Motivated by the formula of the expected number of loops in
the uncorrelated random network[10,20], we derive the expected
number of loops on single nodes in undirected and directed random
networks.

For undirected random networks, the expected number $E(N_{L})$ of
short loops with length L is given by[20]
\begin{equation}
 E(N_{L})=\frac{1}{2L}(\frac{<k(k-1)>}{<k>})^{L},
\end{equation}
where $k$ is the degree sequence of the network and $<.>$ represents
the average value of a sequence. We can obtain the expected number
of short loops on a specific node as
\begin{eqnarray}
E(C_{L}(i))=&
\frac{1}{2L}(\frac{a}{b})^{L}-\frac{1}{2L}(\frac{(a-k_{i}(k_{i}-1))(b-4k_{i})}{(b-2k_{i})^{2}})^{L},
\end{eqnarray}
where $a=\sum\limits_{h=1}\limits^{N}k_{h}(k_{h}-1)$ and
$b=\sum\limits_{h=1}\limits^{N}k_{h}$.

For directed random networks, the expected number $E(N_{L})$ of short
loops with length $L$ can be obtained by[10]
\begin{equation}
 E(N_{L})=\frac{1}{2L}(\frac{<k_{in}k_{out}>}{<k_{in}>})^{L}.
\end{equation}
Like the undirected network, we also deduce a formula to estimate
the expected number of short loops on a specific node $i$. The
formula of $E(C_{L}(i))$ is
\begin{eqnarray}
E(C_{L}(i))=&\frac{1}{L}(\frac{c}{d})^{L}-\frac{1}{L}(\frac{(c-k^{in}_{i}k^{out}_{i})(1-\frac{k^{in}_{i}}{d-k^{out}_{i}}-\frac{k^{out}_{i}}{d-k^{in}_{i}})}{d-k^{in}_{i}-k^{out}_{i}})^{L}.\quad
\end{eqnarray}
where $c=\sum\limits_{h=1}\limits^{N}k^{in}_{h}k^{out}_{h}$ and
$d=\sum\limits_{h=1}\limits^{N}k^{in}_{h}$.  To examine the validity
of our formula, we calculate the exact short loops number[8] in
directed and undirected random networks with prearranged poisson
degree sequences and compare them with the expected values from our
formulas. The result shows our formulas can perfectly predict the
$C_{L}(i)$ of both undirected and directed random networks with
given degree sequences. So far, given the degree sequence of
directed and undirected networks, we can use these two formulas to predict
the loops number on each node in the uncorrelated random networks.
That is to say the distribution $C_{L}^{c}(i)$ can be represented by
the distribution of $E(C_{L}^{c}(i))$ which can be simply calculated
by our formulas.

\begin{figure}
  \center
  \includegraphics[width=8cm]{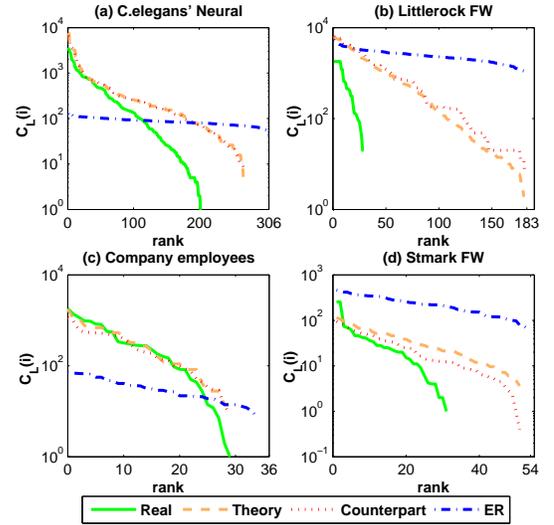}
  \caption{The zipf plots of $C_{L}^{r}(i)$, $C_{L}^{c}(i)$, $C_{L}^{e}(i)$ and $E(C_{L}^{c}(i))$ of four different networks including (a)C.elegans' neural network,
   (b)Littlerock food web network, (c)High technology company employees' friendship network, (d)Stmark food web network.
   In this Figure, we use $L=5$ as example. The links for the random networks are averaged by 100 times.}
\end{figure}

\begin{figure}
  \center
  \includegraphics[width=8cm]{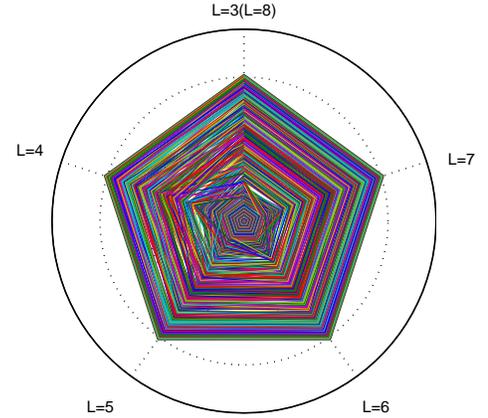}
  \caption{The rank clocks of $C_{L}(i)$ of C.elegans' neural network, with each axis running from rank 1 on the circumference to 306 at the center. The top rank 1 represents the node with the most loops passing through while the bottom rank 306 is the node with the fewest loops.
  Each vertex on the circumference stands for the length of the loops.}
\end{figure}

In fig.1, we compare the zipf plots of $C_{L}^{r}(i)$,
$C_{L}^{c}(i)$, $C_{L}^{e}(i)$ and $E(C_{L}^{c}(i))$ in four
different networks including the C.elegans' neural network, the
littlerock food web network, high technology company employees'
friendship network and the stmark food web network. If a node has no
loop passing through, $C_{L}(i)=0$. In the zipf plot, this node will
not appear in the log axis. Hence, the shorter tail of the line
means all the loops are inclined to locate in several specific nodes. In
addition, the steeper slope of the line in zipf plots indicates the
distribution of $C_{L}(i)$ is more skewed, which means that nodes
are quiet different from each other in loops number. If these two
features are more significant, the distribution of the $C_{L}(i)$
will be more heterogenous. Comparing the real networks to ER random
networks, we find that the loops are more heterogenous in some real
networks such as neural networks and some food web networks.
Moreover, from the $C_{L}^{r}(i)$ and $C_{L}^{c}(i)$, we can easily
find that the degree sequence are not sufficient to describe the
heterogeneity of loops distribution on nodes. In many cases, $C_{L}^{r}(i)$ performs a more significant heterogeneity than
$E(C_{L}^{c}(i))$ and $C_{L}^{c}(i)$. However, loops are distributed in
some social networks almost the same as in the counterpart networks.
A typical example is given in fig.1(c).

In particular, we study $C_{L}(i)$ with different lengths in the
C.elegans' neural networks. In fig.2, we use the rank clocks to test
whether the rank of loops on each node varies dramatically with different loops
length $L$[21]. In fig.2, the rank clocks show perfect pentagon in
the circumference which means that the top ranks of $C_{L}(i)$ do
not change. Hence, we can clearly see that although for different
length $L$, the heterogenous loops distributions share the same top
rank.

From the heterogenous distribution of the loops in each node, we can
easily find that loops locate more uneven in some real networks than
in counterpart random ones. That is to say, some nodes of the real
network are relatively rich in loops while loops are sparse in some
other nodes. This phenomenon indicates that besides the total
number, the detail organization of these loops in the real networks
is quiet different from the counterpart random networks. If these
nodes with many loops tightly connect with each other in the same community, of course this
community will be extremely rich in loops. Combined with the result
in fig.2, the same top rank in the heterogenous loops distribution
will enhance the richness in loops in the community. In the
following section, we will discuss the phenomenon of uneven loops
location by studying the short loops in a specific kind of community
of networks.

\textbf{The rich loop core phenomenon.} To investigate the detail
organization of the self-avoiding loops in a network, we will study
the loops number of different communities. In this paper, we consider
that a loop belongs to a community only if all the nodes of the
loops are included in this community. For these real networks we are
about to analyze in this paper, because long loops are strongly
related to the size of a chosen community, short loops are the main
elements in the community. Therefore, we only consider short loops
in this paper. It has been investigated that many real networks have
fewer loops than the randomized counterparts in both short loops and
long loops such as C.elegans' neural network, Food Web networks,
Power-grid networks etc[10]. Here, we take some of these real
networks to make a further study of the loops in their communities.

\begin{figure}
  \center
  \includegraphics[width=7cm]{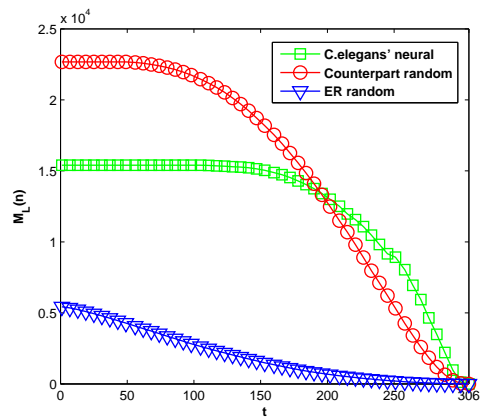}\\
  \caption{The change of the loops number as the nodes removing from three kinds of networks. In this figure, we choose $L=5$ as a example. In fact, because the community obtained in this process is the selected community,
  the loops number in these communities is $M_{L}(n)$. Furthermore, the size of selected community $n=N-t$ in each step $t$. The links for the random networks are averaged by 50 times.}
\end{figure}

\begin{figure}
  \center
  \includegraphics[width=7cm]{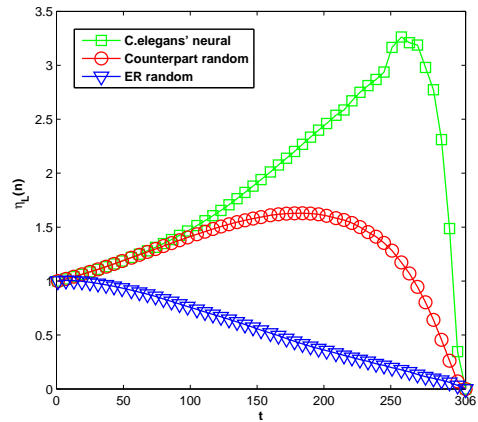}\\
  \caption{The loop density ratio $\eta_{L}(n)$ of three network in each step of node removing process. We choose $L=5$ as a example.
  It is clear that loops locate more unevenly in C.elegans' neural network than the counterpart random networks.
  Furthermore, the size of selected community $n=N-t$ in each step $t$. The links for the random networks are averaged by 50 times.}
\end{figure}

First, we introduce a node removing process. In each step $t$ of the
node removing process, we remove the node with the smallest $C_{L}(i)$
in the network. It means that the node with fewer loops passing
through will be removed first. Specifically, the $C_{L}(i)$ should
be updated in each step. After $N$ steps, we can remove all the nodes
from the network. If we want to obtain a community with size $n$,
the number of the removed nodes should be $t=N-n$. Because the node
about to be removed in every step has the fewest loops among the
remaining nodes, the community obtained by this process has the richest
short loops among all the communities with the same size. The
community obtained by the node removing process is called the selected
community in this paper, the loops number in this community is
denoted as $M_{L}(n)$. Obviously, $M_{L}(N)=N_{L}$ and $M_{L}(0)=0$.
If a network is greatly uneven in the loop location, $M_{L}(n)$ will
decrease slowly in the beginning while decline dramatically in the
end through out the node removing process.

\begin{figure}
  \center
  \includegraphics[width=8cm]{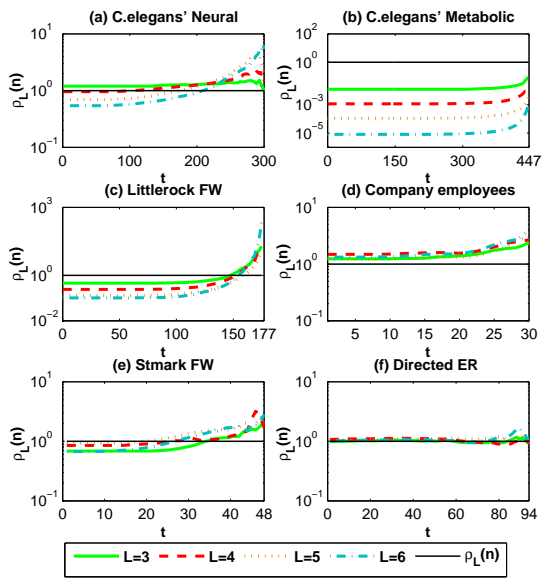}\\
  \caption{Typical results of comparing selected communities' loops of many networks to that of the counterpart random models. The results are averaged by 50 times. (a) C.elegans' neural
  network, (b) C.elegans' metabolic network, (c) and (e) are food web networks in two different places, (d) Company employees' friendship network which is obtained
  by questionnaire, (f) directed ER random networks with 100 nodes and 800 links.}
\end{figure}

We analyze the C.elegans' neural network by the node removing
process as an example. In fig.3, we compare the C.elegans' neural
network, the counterpart random networks and the corresponding ER
random networks. It can be seen that although the counterpart random
networks have more short loops than C.elegans' neural network, it
can not compete with the C.elegans' neural network in some specific
selected communities. For instance, the selected community with 50
nodes in the C.elegans' neural network has more short loops than
that in the counterpart random networks as in fig.3.

Furthermore, we define the loop density as the proportion of the
loops number and the community size. So we can easily get the loop
density of the original network $N_{L}/N$. To compare the loop
density of the selected communities and the original network, we
define the loop density ratio $\eta_{L}(n)$ in each selected
community during the node removing process as

\begin{equation}
\eta_{L}(n)=\frac{M_{L}(n)N}{N_{L}n}.
\end{equation}

The loop density ratio $\eta_{L}(n)$ is the proportion of the
selected community's loop density and the original network's loop
density. If the loops locate unevenly in a network, $\eta_{L}(n)$
will become larger than $1$ during the node removing process. It
indicates some communities in the network have bigger loop density
than the original network. Obviously, in the C.elegans' neural
network, some of the selected communities enjoy larger loop density
than the original network, so $\eta_{L}(n)$ is larger than $1$ as
shown in Fig.4. Although the counterpart random networks can also
have a $\eta_{L}(n)$ larger than $1$, its value is always lower than
that of C.elegans' neural network. This feature indicates that loops
locate more unevenly in C.elegans' neural network than in the
counterpart random networks, which means that lots of loops are
limited in some specific nodes of the C.elegans' neural network.

So the unevenness may result in a rich loop core phenomenon, which
means some selected communities with far higher loop density
compared with the original networks. Although many real networks
have fewer short loops compared to counterpart random models, the
rich loop phenomenon will make some communities in real networks
more loopy than the corresponding communities in the counterpart
random networks. It can be detected by the ratio as:
\begin{equation}
\rho_{L}(n)=\frac{M_{L}^{real}(n)}{M_{L}^{rand}(n)},
\end{equation}
where $n$ is the size of the selected community. For example, the
index will turn from $\rho_{L}(n)<1$ to $\rho_{L}(n)>1$ as nodes
removed in the C.elegans' neural network as show in fig.3. Also, we
also investigate many other real networks, some typical results are
shown in fig.4.

As in fig.4 (a), (c) and (e), although these real networks, such as
neural networks and some food web networks, have fewer total loops
than the counterpart random networks, some selected communities of
them are more loopy than that of counterpart random ones. Moreover,
we find that short loops with different lengths $L$ perform the same
trend in the rich loop core phenomenon. If the loops with specific
length $L$ locate unevenly in the networks, loops with other lengths
locate uneven as well. However, not all the networks have this kind
of phenomenon as shown in Fig.4(b). Some real networks with far
fewer loops than the counterpart random networks have
$\rho_{L}(n)<1$ for all $n$. For example, the C.elegans' metabolic
network belongs to this category. In addition, we find that some
social networks have more short loops than the counterpart random
ones as in fig.4(d). These networks are not discussed in ref[10].
This category includes the prisoners' friendship network, high
technology employees' friendship network, the family visit network,
the flying-team partner choosing network, the dining table partner
choosing network and so on[22]. If these networks enjoy more uneven
loops location, their selected communities can only be more and more
loopy than that of counterpart random networks as the community size
$n$ varies, see fig.4(d). Finally, we use the index $\rho_{L}(n)$ to
detect a ER random network with 100 nodes and 800 links. The result
in fig.4(f) shows that $\rho_{L}(n)=1$ approximately.

Besides the rich loop core phenomenon, how to find the rich loop
core in a network is an interesting question. In this paper, we
simply consider the rich loop core appears at the maximum
$\eta_{L}(n)$, which means the rich loop core will have highest
loop density than any other community in the networks. Some typical
results are shown in fig.6. The littlerock food web and the
C.elegans' metabolic network have significant rich loop cores which
indicates they have much higher loop density community compared with the
original networks. The C.elegans' neural networks and the StMark
food web have such rich loop cores as well. On the contrary, this
phenomenon is not so obvious in company employees networks and large
degree ER random networks. For the C.elegans' metabolic network, the
high loop density ratio is due to the small number of total loops.
These loops unavoidably locate in several specific nodes, so the
loops density in the selected community will be very large compared
with the original network. Furthermore, we analyze the rich loop
cores of the food web and neural networks. It is interesting that
most of the nodes in the cores are from interneurons in C.elegans'
neural network and from microorganisms in food web networks.
Specifically, the rich loop cores of C.elegans' neural networks have
different sizes $n$ for different loop lengths $L$. For example,
$n=31$ when $L=3$, $n=38$ when $L=4$, $n=50$ when $L=5$ and $n=51$
when $L=6$. These rich loop cores share $30$ nodes and $24$ of them
are interneurons. The C.elegans' neural networks are composed by
sensory neurons, interneurons and motor neurons. Most interneurons
are in the nerve ring ganglia. Their main function is to process
signals[23,24]. So the loops number in these interneurons is
relatively larger than the others. Likewise, the rich loop cores in
stmark food web networks are $n=20$ when $L=3$, $n=22$ when $L=4$,
$n=24$ when $L=5$ and $n=24$ when $L=6$. These rich loop cores share
$14$ nodes and $10$ of them are microorganisms. In the food web
networks, the microorganisms are the main element in microbial loops
which are strongly related to fixed carbon repacking and recovery
path of ecosystem[19].

\begin{figure}
  \center
  \includegraphics[width=8cm]{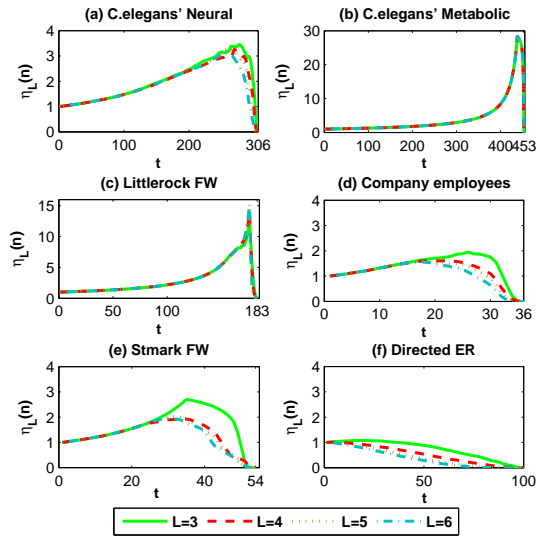}\\
  \caption{The rich loop core phenomenon of some typical real networks. The links for the random networks are averaged by 50 times. (a) C.elegans' neural
  network, (b) C.elegans' metabolic network, (c) and (e) are food web networks in two different places, (d) Company employees' friendship network which is obtained
  by questionnaire, (f) directed ER random networks with 100 nodes and 800 links.}
\end{figure}

\textbf{Measurement for unevenness of loop location.} In order to
quantify how unevenly loops locate in the networks, we present an
index which bases on the node removing process. In order to Simplify
the computing complexity, we do not update $C_{L}(i)$ in each step
during the node removing process in this section. We test and find
the result obtained in way is sufficient to represent that by
updating $C_{L}(i)$ in each step statistically. Moreover, we remove the nodes based on
the attacking rate $p$. For example, if $p=0$, no node is removed
and the network is all the same with the original network and the loop number is
$N_{L}$. If $p=0.1$, we just remove $[pN]_{ceil}$ nodes from the
network and the loop number is $M_{L}(N-[pN]_{ceil})$. Here,
$[.]_{ceil}$ represents the operation of rounding upward. Then, we
use $A_{L}(p)=\frac{M_{L}(N-[pN]_{ceil})}{N_{L}}$ to normalize the
loop number of each community so that $A_{L}(p)$ which is
corresponding to the loop number declines from 1 to 0 through out
the node removing process as shown in fig.7. Again, we use $L=5$ as
an example in fig.7.

If a network is significantly uneven in the loop
location, $A_{L}(p)$ will decline slowly in the beginning while
dramatically in the end during the node removing process. On the contrary, if loops locate evenly in the network, the $A_{L}(p)$
 will decline almost the same as the corresponding ER random networks. Therefore, the unevenness of loop location can be measured by the
difference between the real network and the corresponding ER random
network. Here, we use $I_{L}^{r}(p)=A_{L}^{r}(p)-A_{L}^{e}(p)$ to
estimate the difference. So the unevenness of loops location can be
represented by the index as
\begin{equation}
 R_{L}=\int_{0}^{1} I_{L}(p)dp,
\end{equation}
where $-1<R_{L}<1$. The severer the unevenness is, the larger
the index $R_{L}$ is, which means the loop location departs more
largely from the corresponding ER random network. Of course, the
index $R_{L}$ can also be used in analyzing the counterpart random
network.

Actually, the index $R_{L}$ is the area between the lines of the
real network and the ER random network in fig.7. It can be seen that
the real networks and the counterpart random networks can be different in the
unevenness, as the index $R_{L}^{r}\neq R _{L}^{c}$. Typically, if
the real network is very sparse in the links, the network has only
small number of loops. This will lead to a phenomenon that some part
of the network has some loops, while other part has no loop at all.
However, the corresponding ER random network has the same condition
too. For the $R_{L}$ is the area between the lines of the real
network and the ER random network, the $R_{L}$ will be a small value
under this circumstance. Whether this uneven location of loops results from the specific structure of the real network or from the
small degree as the ER random network can be estimated from the
index $R_{L}$. As the figure(d) in the fig.7, we use the index
$R_{L}$ to investigate a ER random network with small degree as
$K=2$. The result shows that the three lines are almost the same
meaning that $R_{L}\approx0$.

\begin{figure}
  \center
  \includegraphics[width=4cm]{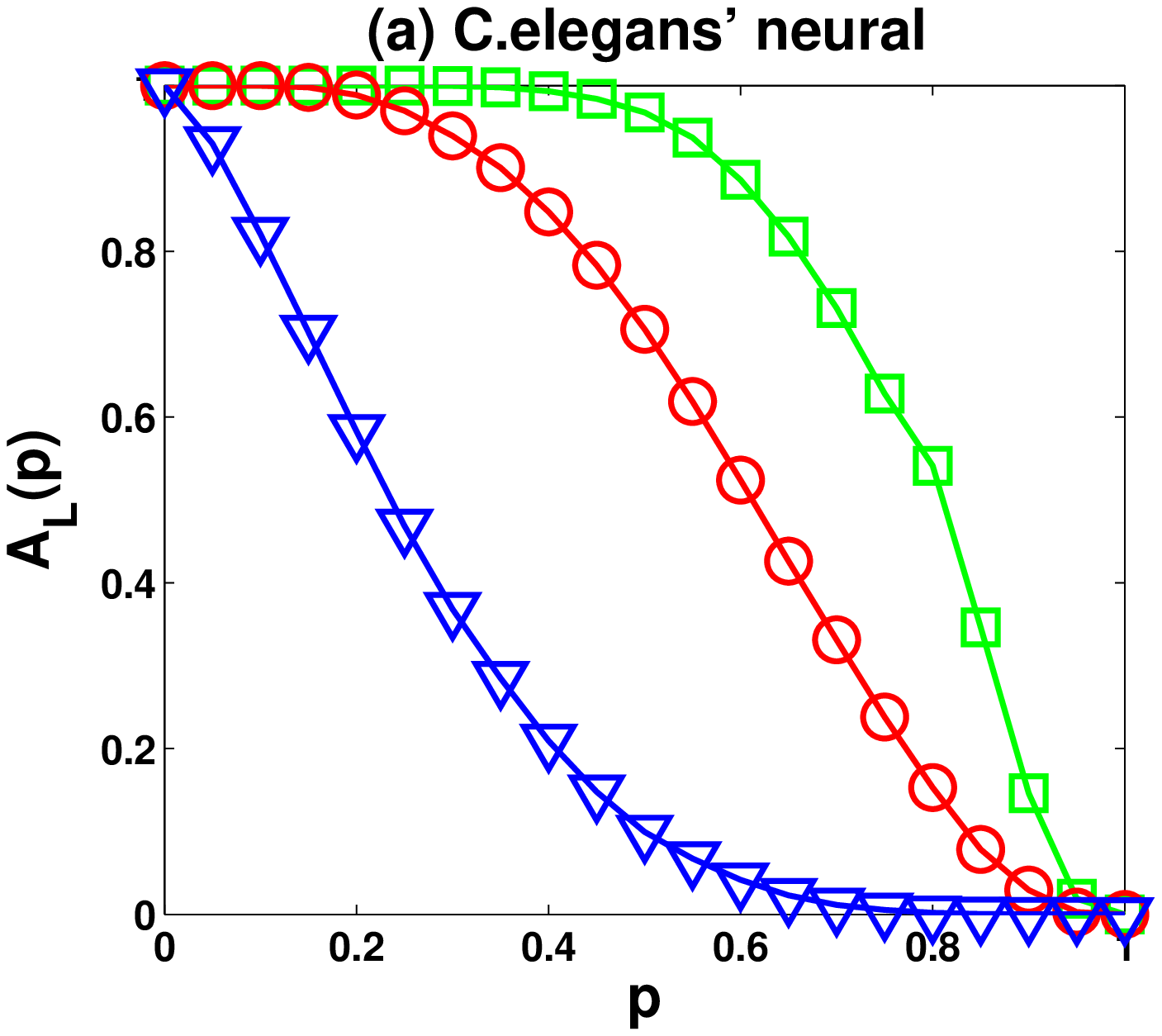} \includegraphics[width=4cm]{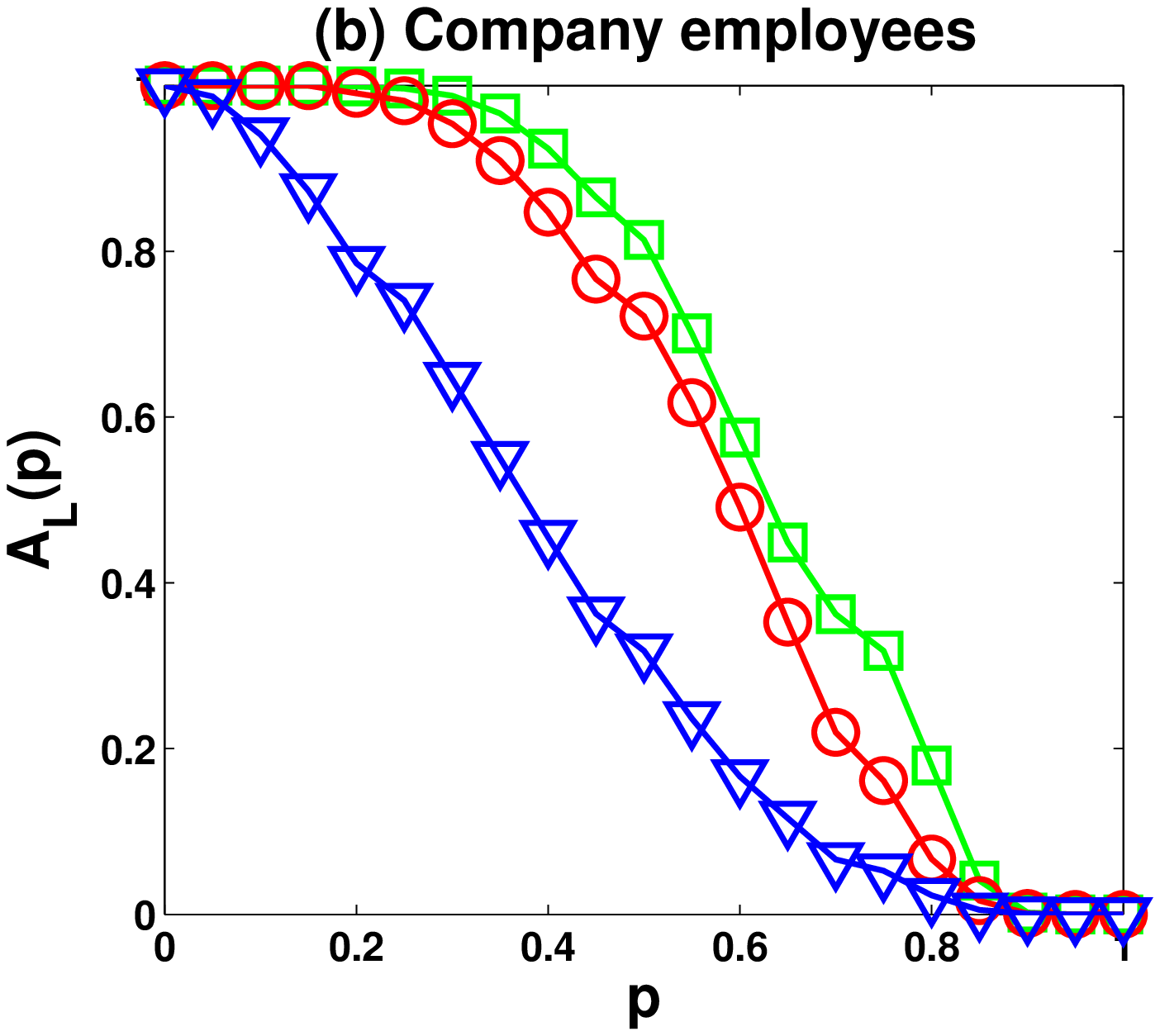}\\
  \includegraphics[width=4cm]{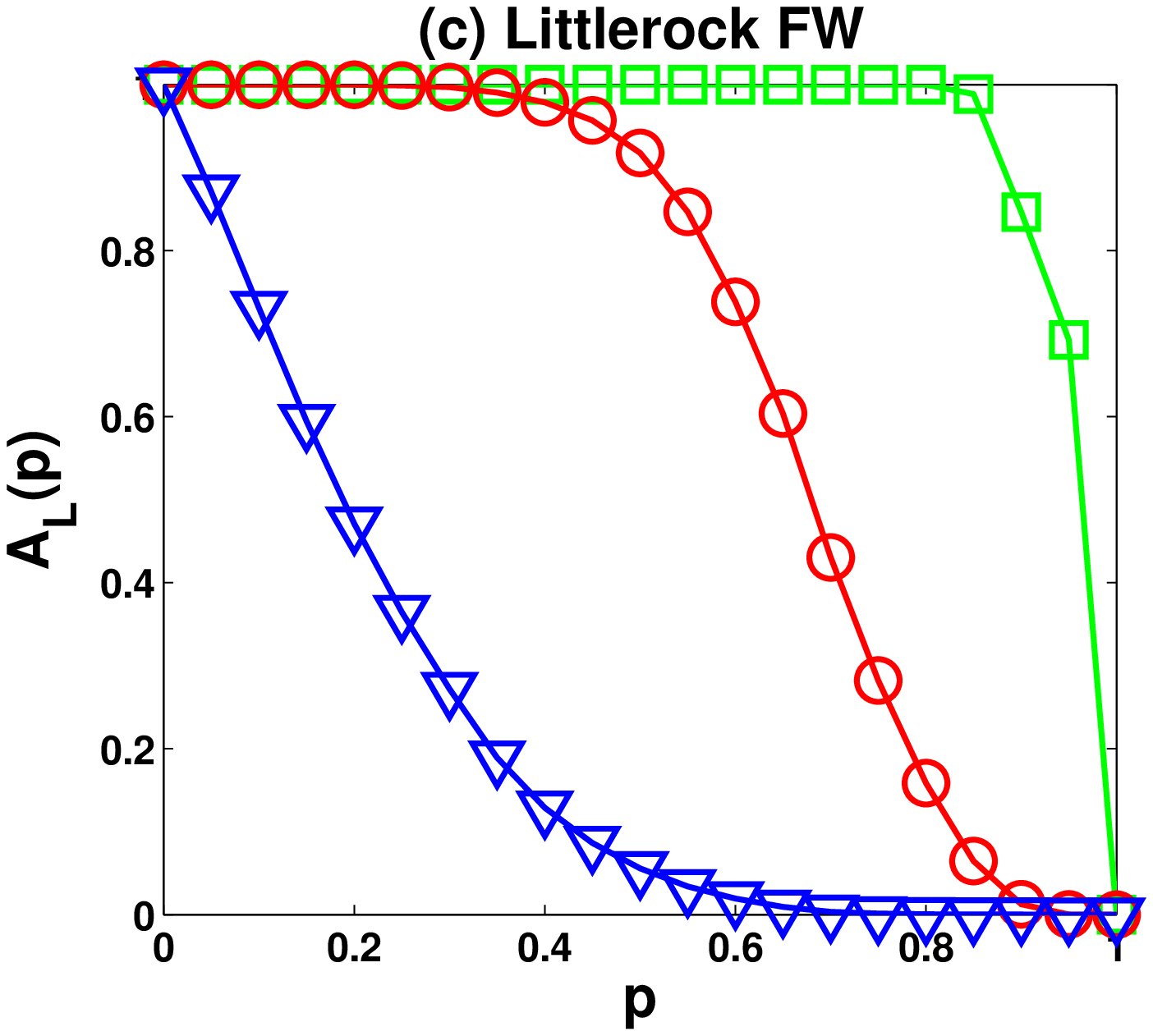} \includegraphics[width=4cm]{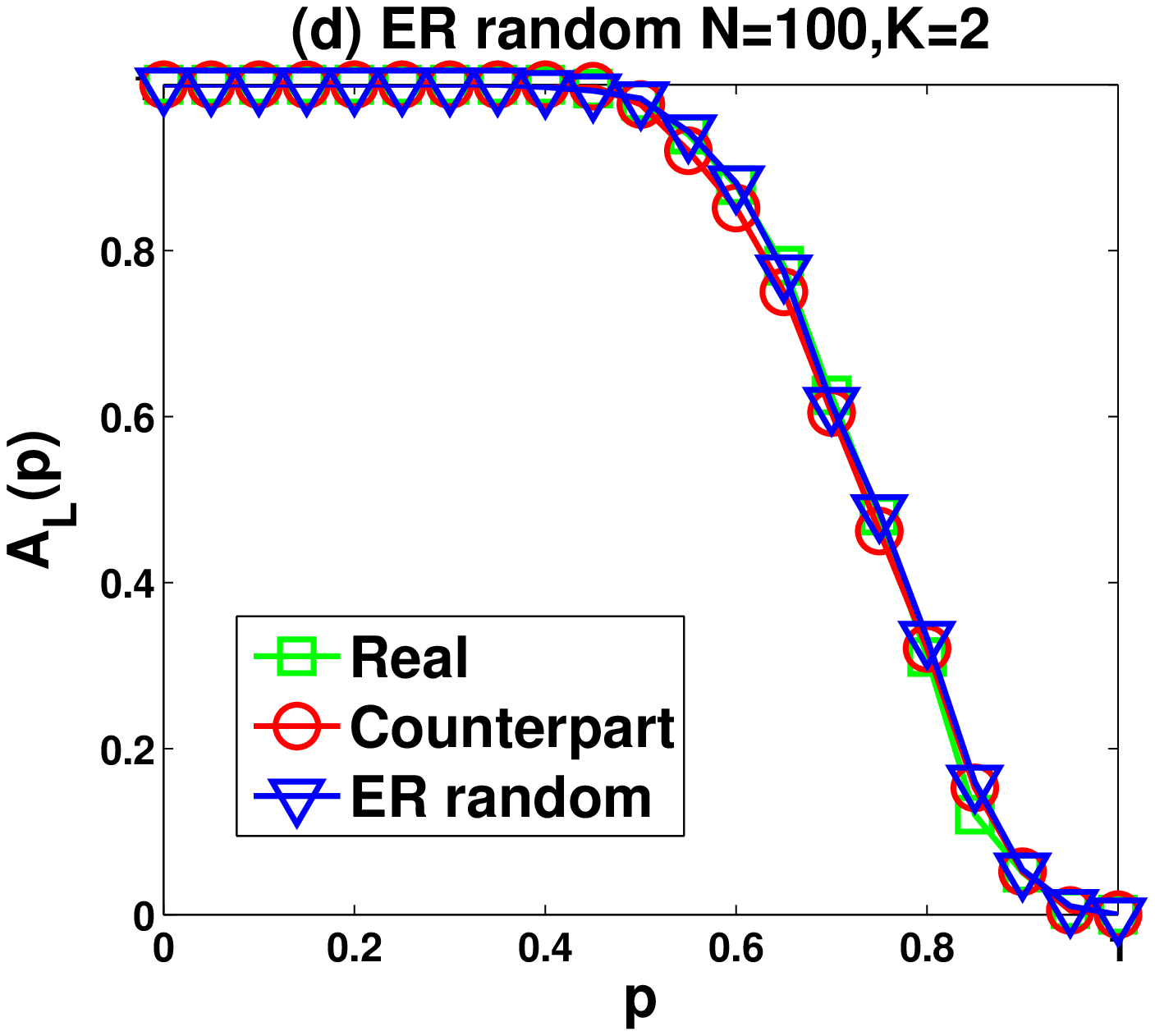}\\
  \caption{the $A_{L}^{r}(p)$, $A_{L}^{c}(p)$ and $A_{L}^{e}(p)$ change as the $p$ in different networks including C.elegans' neural network, high technology company employees' friendship network, littlerock FW network, and ER random network with small degree.
   $L$ is chosen as 5 and the result is averaged by 50 times.}
\end{figure}

\begin{table}[!htb]
 \extrarowheight=0.4em
 \tabcolsep=4pt
\begin{center}
\caption{Results of the analysis of networks based on index
$\bar{R}$}
\begin{tabular}{lccclccc}

\hline
  network  &$size$ &$links$  &$\bar{R}_{r}$ &$\bar{R}_{c}$ &$\bar{R}_{r}-\bar{R}_{c}$\\

\hline
  C.elegans' neural &$306$ &$2359$ & $0.436$ & $0.317$ &$0.119$\\
  C.elegans' metabolic &$453$ &$2040$ & $0.468$ &$0.337$ &$0.131$\\
  E.coli's metabolic &$896$ &$958$ &$-0.021$ &$-0.002$ &$-0.019$\\
  Mondego FW  &$46$ &$400$ &$0.231$ & $0.257$ &$-0.026$\\
  Michigan FW &$39$ &$221$ &$0.194$ & $0.207$ &$-0.013$\\
  Littlerock FW &$183$ &$2494$ &$0.672$ &$0.439$ &$0.233$\\
  StMarks FW  &$54$ &$356$ &$0.378$ &$0.256$ &$0.122$\\
  Prisoners   &$67$ &$182$ &$0.169$ &$0.004$  &$0.165$\\
  Flying-teamers &$48$ &$351$ &$0.108$ &$0.060$  &$0.048$\\
  Company employees &$36$ &$147$ &$0.219$ &$0.163$  &$0.056$\\
  ER random  &$--$ &$--$  &$0$ \quad\quad  &$0$  &$0$\quad\quad\\

 \hline
\end{tabular}
\end{center}
\end{table}

Additionally, we consider several more directed real networks[22].
As mentioned above, although each network has different kinds of
short loops based on length $L$, these loops perform almost the same.
We use average $R_{L}$ to represent the unevenness in loops
location. It can be gained by $\bar{R}=<R_{L}>$ and the
$L=3,4,...,L_{max}$ where $L_{max}=8$ according to Ref[9]. The index
$R$ for these real networks are given in table 1. From the
$\bar{R}_{r}$, how significant the unevenness in loops location is
can be known. By comparing the $\bar{R}_{r}$ and the $\bar{R}_{c}$,
we can distinguish whether this unevenness results from the degree
sequence. In table 1, it can be seen from the $\bar{R}_{r}$ that the
C.elegans neural network, C.elegans metabolic network and some food web networks are
really uneven in loop location. The social networks, Escherichia Coli's
metabolic network and some other food web networks do not have such significant unevenness. In fact,
the number of different species will affect the loop location in
food web networks. For example, too many microorganisms will make
the loops more even and too many metazoans will reduce the total
loops number, at this time the rich loop core will be more obvious. Additionally,
both the C.elegans' metabolic network and the Escherichia Coli's
metabolic network have very few loops, but the degree sequence and
the total links of the former one allows the counterpart random and
the ER random networks to have much more loops while the latter one
does not. Hence, the $\bar{R}_{r}$ of these two metabolic networks
are different. Moreover, comparing the $\bar{R}_{r}$ and the
$\bar{R}_{c}$ in table 1, it can be found that the degree sequence
is not sufficient to describe the unevenness in loops location. It
is clear that the C.elegans' neural network, littlerock and stmark
food web networks are more uneven in loops location than the
counterpart networks. That is why they can have some communities
more loopy than the corresponding communities in the counterpart
random models despite the fewer total loops number.

\textbf{Conclusion.} The previous works on the loops mainly focus on
the total number of loops and the dynamic effect of the loop
structure. However, the loop location is also very important in
networks. Generally, loops tend to locate in some specific nodes in
some real networks, which means some communities of the network are
extremely rich in loops while the loops are relatively sparse in
other parts. If this uneven location is significant enough, the rich
loop core phenomenon can be formed in some real networks.

The rich loop core phenomenon is meaningful for the typical
function of real networks. For instance, the loop structure is
strongly related to the self-sustained activities in neural networks,
so the rich loop core may help to understand the functional regions
in the neural networks. For the food web networks, almost all the
loops in rich loop cores of food web networks are microbial loops and
plays an important part in fixed carbon repacking and recovery path
of ecosystem. In addition, this uneven location of loops may provide
a new way to study the community detection in directed networks,
which asks for further research.

\textbf{Acknowledgement.} The authors would like to thank Dong Zhou,
Hongzhi You, Prof. Ying Fan and Prof. Yiming Ding for many useful
suggestions. This work is supported by NSFC under Grants No.
70974084, No. 60534080 and No. 70771011.


\begin{thebibliography}{99}

\bibitem{1} R. Pastor-Satorras, A. Vazquez, A. Vespignani, Phys. Rev. Lett. 87, 258701 (2001).
\bibitem{2} S. N. Dorogovtsev et al., Rev. Mod. Phys. 80, 1275 (2008).
\bibitem{3} B. Karrer and M. E. J. Newman, Phys. Rev. Lett. 102, 128701 (2009)
\bibitem{4} A.Zeng, Y.Hu and Z.Di, Europhys. Lett. 87, 48002 (2009)
\bibitem{5} K. Klemm and P. F. Stadler, Phys. Rev. E 73, 025101(R) (2006)
\bibitem{6} E. Marinari, G. Semerjian and V. VanKerrebroeck, Phys. Rev. E 75, 066708 (2007)
\bibitem{7} E. Marinari, R. Monasson and G. Semerjian, Europhys. Lett. 73, 8-14 (2006)
\bibitem{8} G. Bianconi, Eur. Phys. J. B 38, 223¨C230 (2004)
\bibitem{9} G. Bianconi and N. Gulbahce, J. Phys. A: Math. Theor. 41, 224003 (2008)
\bibitem{10} G. Bianconi, N. Gulbahce and A. E. Motter, Phys. Rev. Lett. 100, 118701 (2008)
\bibitem{11} H. D. Rozenfeld et al., J. Phys. A: Math. Gen. 38, 4589¨C4595 (2005)
\bibitem{12} A. Roxin, H. Riecke and S. A. Solla, Phys. Rev. Lett. 92, 198101 (2004)
\bibitem{13} X. Liao et al., arXiv:0906.2356v1
\bibitem{14} L. F. Lago-Fernandez, R. Huerta, F. Corbacho, J. A. Siguenza, Phys. Rev. Lett. 84, 2758-2761 (2000)
\bibitem{15} X. Ma,L. Huang, Y. C. Lai and Z. Zheng, Phys. Rev. E 79, 056106 (2009)
\bibitem{16} A. Arenas, A. D¨ªaz-Guilera, J. Kurths, Y. Moreno and C. Zhou, Physics Reports 469,  93-153 (2008)
\bibitem{17} J.D. Noh, Eur. Phys. J. B 66, 251-257 (2008)
\bibitem{18} V. VanKerrebroeck and E. Marinari, Phys. Rev. Lett. 101, 098701 (2008)
\bibitem{19} E. Sherr and B. Sherr, Limnology and Oceanography, 33, 1225-1227 (1988)
\bibitem{20} Z. Burda, J. Jurkiewicz, A. Krzywicki, Phys. Rev. E 70, 026106 (2004)
\bibitem{21} M. Batty, Nature 444, 592-596 (2006)
\bibitem{22} The network datas are available on line at http://vlado.fmf.unilj.si/pub/networks/data/, http://www.imsc.res.in/~sitabhra/research/neural/celegans/, http://www.casos.cs.cmu.edu/index.php, www.cosinproject.org/.
\bibitem{23} J. G. White, E. Southgate, J. N. Thomson and S. Brenner, Philos Trans R Soc London B, Biol Sci 314, 1-340 (1986)
\bibitem{24} E. L. Tsalik, O. Hobert, J Neurobiol 56(2),178-97 (2003)

\end{thebibliography}
\end{document}